\documentclass[aps,prd,twocolumn,superscriptaddress,groupedaddress]{revtex4}  % for review and submission
\pdfoutput=1
\usepackage{graphicx}  
\usepackage{dcolumn}   
\usepackage{bm}       
\usepackage{amssymb}   
\usepackage{amsmath}
\usepackage{lipsum}
\usepackage{placeins}
\usepackage{enumitem}
\usepackage{hyperref}

\usepackage{array}
\newcolumntype{C}[1]{>{\centering\let\newline\\\arraybackslash\hspace{0pt}}m{#1}}
\newcolumntype{L}[1]{>{\raggedright\let\newline\\\arraybackslash\hspace{0pt}}m{#1}}
\newcolumntype{R}[1]{>{\raggedleft\let\newline\\\arraybackslash\hspace{0pt}}m{#1}}

\def\ieec{Institut de Ci\`encies de l'Espai, (CSIC-IEEC),
Campus UAB,  Facultat de Ci\`encies, Torre C-5, 08193 Bellaterra, Spain}
\def\unitn{Dipartimento di Fisica, Universit\`a di Trento and INFN,
Gruppo Collegato di Trento, 38123 Povo, Trento, Italy}
\def\esac{ESAC, European Space Agency, Camino bajo del Castillo s/n,
Urbanizaci\'on Villafranca del Castillo, Villanueva de la Can\~ada, 28692 Madrid, Spain}
\def\aei{Albert-Einstein-Institut, Max-Planck-Institut f\"ur Gravitationsphysik und Universit\"at Hannover,
Callinstrasse 38, 30167 Hannover, Germany}
\def\apc{APC, Universit\'e Paris Diderot, CNRS/IN2P3, CEA/Ifru, Observatoire de Paris,
Sorbonne Paris Cit\'e, 10 Rue A.\;Domon et L.\;Duquet, 75205 Paris Cedex 13, France}
\def\estec{European Space Technology Centre, European Space Agency, Keplerlaan 1, 2200 AG Noordwijk, The Netherlands}
\def\ethz{ETH Z\"urich, Institut f\"ur Geophysik, Sonneggstrasse 5, 8092 Z\"urich}

\begin{document}

\title{Bayesian Model Selection for LISA Pathfinder}

\author{Nikolaos~Karnesis}
\email{karnesis@ieec.uab.es}
\author{Miquel~Nofrarias}
\author{Carlos F.~Sopuerta}\author{Ferran~Gibert}\affiliation{\ieec}
\author{Michele~Armano}\affiliation{\esac}
\author{Heather~Audley}\affiliation{\aei}
\author{Giuseppe~Congedo}\affiliation{Department of Physics, University of Oxford, Keble Road, Oxford OX1 3RH, UK}
\author{Ingo~Diepholz}\affiliation{\aei}
\author{Luigi~Ferraioli}\affiliation{\ethz}
\author{Martin~Hewitson}\affiliation{\aei}
\author{Mauro~Hueller}\affiliation{\unitn}
\author{Natalia~Korsakova}\affiliation{\aei}
\author{P.W.~McNamara}\affiliation{\estec}
\author{Eric~Plagnol}\affiliation{\apc}
\author{Stefano~Vitale}\affiliation{\unitn}

\vskip 0.25cm

\date{\today}

\begin{abstract}

The main goal of the LISA Pathfinder (LPF) mission is to fully characterize the acceleration noise models and to test 
key technologies for future space-based gravitational-wave observatories similar to the eLISA concept. The data analysis team has 
developed complex three-dimensional models of the LISA Technology Package (LTP) experiment on-board LPF.
 These models are used for simulations, but more importantly, they will be 
 used for parameter estimation purposes during flight operations. One of the tasks of the data analysis team is to 
 identify the physical effects that contribute significantly to the properties of the 
 instrument noise. A way of approaching this problem is to recover the essential parameters of a LTP model fitting the
  data. Thus, we want to define the simplest model that efficiently explains the observations.
To do so, adopting a Bayesian framework, one has to estimate the so-called 
Bayes Factor between two competing models. In our analysis, we use three main different methods to estimate it: 
the Reversible Jump Markov Chain Monte Carlo method, 
the Schwarz criterion, and the Laplace approximation. They are applied to simulated LPF experiments where
the most probable LTP model that explains the observations is recovered. The same type of analysis presented in this paper is expected to 
be followed during flight operations. Moreover, the correlation of the 
output of the aforementioned methods with the design of the experiment is explored.

\end{abstract}

\maketitle

%
%
%
% INTRODUCTION
%
%
%

\section{\label{sec:level1}Introduction}

LISA Pathfinder (LPF) \cite{lpf:2012zz,macnamara2013} is an European Space Agency mission that will serve as a technology 
demonstrator for a future space-based gravitational-wave observatory like eLISA~\cite{whitepaper}. The LPF mission will prove geodesic motion by monitoring 
the relative acceleration of two test masses in nominally free-fall in the frequency band of 1 to 30 mHz.
 The main instrument on-board LPF is the LISA Technology Package (LTP), which is a suite of experiments with the aim of measuring and characterizing the
different contributions to the differential acceleration noise between
the two test masses.  This characterization is the main task of the data analysis team.
To that end, dedicated experiments are going to be performed in order to estimate the unknown parameters 
 of the system. And for that purpose, a number of parameter estimation methods and models of the LTP have been 
 implemented \cite{Nofrarias2010,Congedoprd,Antonucci2011}. The main question that arises is about the suitability of the different models implemented, or in simpler terms,
  which model can describe better the observations of the experiment.

The motivation to implement an algorithm that will help us classify our models is due to the nature of the LTP system. 
There has been a lot of work trying to understand the instrument, and the models implemented are based on theoretical 
and experimental measurements from test campaigns \cite{felipe,Audley}. 
Selecting the most probable model is crucial for the analysis for two main 
reasons: first, the most suitable LTP model that describes the observed 
physical effects is chosen, and secondly, over-fitting situations are avoided, 
together with biased estimation of the parameters of the system.  

One could use several criteria and algorithms that classify competing models, but in the end,
 working in a Bayesian framework, the main aim is to calculate the {\it Bayes Factor}, which is 
 a comparison between  the {\em evidences} of the models \cite{bf1995,mackay}. The 
 evidence is defined as the probability of the data $\vec{\bf y}$ given the 
 model, that is, the probability distribution on $\vec{\bf y}$ that quantifies the predictive capabilities of the given model.

Most of the approaches are based upon the likelihood evaluated at the maximum and a penalty for the 
number of parameters in the model consisting in multiplying by the so-called {\it Occam Factor}. The {\it Occam's razor} 
is the principle that states that the simplest hypothesis that explains the observations is the most favorable. 
In our case we assume that we have to compare two different LTP models; model $M_1$ and a simpler one, $M_2$, over a data set $\vec{\bf{y}}$. 
If $M_1$ is a more complex model, it presents more predictive capabilities than 
$M_2$, which translates to more disperse evidence over the data set $\vec{\bf y}$. Thus, in the case where the data is compatible with 
 both models, $M_2$ will turn out to be more probable than $M_1$, without having to express any 
 subjective dislike for complex models \cite{mackay}. In other words, it is almost certain that the more parameters 
 in a model the better the fit to the data. But taking it to the extreme, we can imagine a model with as many 
 parameters as the data to fit. In that case we have over-parameterized the model while the aim is to include 
 only the parameters which {\it substantially} improve it. 

In this paper we investigate several methods to compare competing LTP models giving emphasis to Reversible
 Jump Markov Chain Monte Carlo techniques. The plan of the paper is as follows: In Sec.~\ref{sec:background}, we make a brief introduction 
 to Bayesian methods and
 in Sec.~\ref{sec:ltpmodel}, the LTP experiments and models 
 are thoroughly explained and we investigate several applications of our Bayesian techniques to LTP experiments.
  In Sec.~\ref{sec:BFvsExp}, we compare the available methods and discuss their output in connection with the experiment design. 
  We end with a summary and conclusions in Sec. \ref{sec:conclusions}. The codes for the different algorithms are integrated into 
 the LISA Technology Package Data Analysis (LTPDA) Matlab toolbox \cite{Hewitson_ltpda,website:LTPDA}, 
 that comes together with proper documentation and help for the user.
 
%
%
%
% BACKGROUND IN BAYESIAN STATISTICS
%
%
%

\section{\label{sec:background}Background on Bayesian Statistics}

An algorithm that automatically penalizes higher-dimensional models is the Reverse Jump Markov Chain Monte Carlo (RJMCMC) algorithm. The RJMCMC method~\cite{Green1995,Lopes2004,Dellaportas2002, V1, V2} is widely used when dealing 
with nested models, meaning that we need to compare a set of models, where simpler models are a subset of a 
more complicated one. In fact, the RJMCMC algorithm is the generalized case of Markov Chain Monte Carlo (MCMC) methods that is capable of sampling the 
parameter space and at the same time jumping between models with different dimensionality. 
 
 At this point it would be convenient to describe the philosophy of the Bayesian framework and the Metropolis algorithm and then move to the general case of the RJMCMC algorithm. The Bayes 
 rule can be summarized by the following equation:
\begin{equation}
\label{eq:bayes}
\pi (\vec{\bf{\theta}} | \vec{\bf{y}}) = \frac{\pi(\vec{\bf{y}} | \vec{\bf{\theta}})p(\vec{\bf{\theta}})}{\pi(\vec{\bf{y}})}, 
\end{equation}
where $\vec{\bf{\theta}}$ is a model parameter vector,  $\pi(\vec{\bf{y}} | \vec{\bf{\theta}})$ is the {\it likelihood} of the 
parameters $\vec{\bf{\theta}}$ over the data-set $\vec{{\bf y}}$, and $\pi (\vec{\bf{\theta}} | \vec{\bf{y}})$ and 
$p(\vec{\bf{\theta}})$ are the {\it posterior} and the {\it prior} distributions of the parameters respectively. Note that the {\it evidence} 
$\pi(\vec{\bf{y}})$, or {\em marginal likelihood}, is often neglected in parameter estimation algorithms, as it serves 
only as a normalisation constant:
\begin{equation}
\label{eq:simplebayes}
\pi (\vec{\bf{\theta}} | \vec{\bf{y}}) \propto \pi(\vec{\bf{y}} | \vec{\bf{\theta}})p(\vec{\bf{\theta}}). 
\end{equation}

The Metropolis algorithm is one of the MCMC-type methods available that is widely used for parameter estimation
 purposes. It is based on sampling the parameter space by proposing new 
samples $\vec{\theta}_n$ and evaluating the likelihood at each step. By sampling the posterior distribution, probability distributions to the parameters to be estimated are assigned. It is certain that given a large amount of steps in the parameter space, the Metropolis algorithm will converge to the set of parameters 
$\vec{\theta}_{\rm MAP}$ that maximise the likelihood. 

\subsection{\label{sec:calcbayes}Calculating the Bayes Factor}

The evidence of a 
hypothesis ${\rm X}$ given the data-set $\vec{\bf{y}}$, $\pi_{\rm X}(\vec{\bf{y}})$, states the support for this hypothesis, or in other words, how much the 
data favors a given model. In our case, the hypothesis is the model implemented to describe the LTP system. 
Consequently, given two different models X and Y, the Bayes Factor $B_{\rm XY}$ is a comparison between the evidences of model ${\rm X}$ 
and model ${\rm Y}$ given by their ratio
\begin{equation}
B_{\rm XY} = \frac{\pi_{\rm X}(\vec{\bf{y}})}{\pi_{\rm Y}(\vec{\bf{y}})},
\end{equation}
where
\begin{equation} 
\label{eq:evidence}
\pi_{\rm K}(\vec{\bf{y}}) = \int  \pi(\vec{\theta}_{\rm K} , \vec{\bf{y}})d\vec{\theta}, \quad {\rm K = X, Y 
}.
\end{equation}
This integral is extremely costly to evaluate, specially when the model becomes complicated 
with higher dimensionality. In the next sections, a selection of estimators of the Bayes Factor are described. Most of 
them evaluate directly the Bayes factor, while other techniques are used to estimate the evidence for each model. 
All of them are used to investigate our LPF models and mission experiments. In the end, we can draw conclusions
 about the competing models given the estimated value of the Bayes Factor. If $\rm B_{\rm XY} < 1$, the evidence is negative and the observations support model Y. If  
 $\rm B_{\rm XY} > 1$, the evidence is positive and model X is more favourable than model Y.
 
\subsection{The Reversible Jump Markov Chain Monte Carlo Algorithm}
\label{sec:rj}

The RJMCMC algorithm is a robust and efficient tool to estimate the Bayes Factor. It can be shown \cite{Green1995,Dellaportas2002} 
that after a large number of iterations it will converge to the true value of the Bayes Factor.  The only drawback 
is the computational cost of the algorithm. When more than three models are being compared, meaning that 
many transdimensional moves have to be performed, a considerable amount of time is required for convergence. 
The algorithm implemented in this work is a special case of the {\em Metropolized Carlin and Chib} method \cite{Lopes2004}. 
More specifically, let us suppose that we have a total number K of models to compare given a data set $\vec{\bf y}$. Then, the recipe for our RJMCMC method can be 
  summarised in the following steps:
\begin{enumerate}
 \item Initialization: Choose an initial model $k$ and the corresponding parameters $\vec{\theta}_k$.
 
 \item Apply the Metropolis algorithm for model $k$. This step is also called the ``in model step''.
 
 \item Generate new $\vec{\theta_{k'}}$ from a multivariate Gaussian PDF and a random number $\rho$ $\epsilon [0,1]$ from a uniform distribution. This is the 
 step where we propose a new model $k'$. 
  
 \item Calculate the acceptance ratio $\alpha'$:
 
 \begin{equation}
 \label{eq:a'}
 \alpha' = {\rm min}\left[\frac{\pi(\vec{\bf y}|\vec{\theta}_{k'})p(\vec{\theta}_{k'})g(u_{k'})}{\pi(\vec{\bf y}|\vec{\theta}_k)p(\vec{\theta}_k)g(u_k)}|{\bf J}| , 1\right] ,
 \end{equation} 
 
 where $g(u)$ is the proposal distribution from where the ``dimension matching'' parameters $u$ are drawn \cite{Godsill}, and $|{\bf J}|$ is the Jacobian:
 
 \begin{equation}
 \label{eq:j}
 |{\bf J}| = \left|\frac{\partial (\vec{\theta}_{k'},u_{k'})}{\partial 
 (\vec{\theta}_{k},u_{k})}\right|.
 \end{equation} 
 
 \item If $\rho < \alpha'$ we accept the new model $k'$ with parameters $\vec{\theta_{k'}}$ and set $\vec{\theta_k} = \vec{\theta_{k'}}$. 
 
 \item Iterate from step 2 until convergence.
 \end{enumerate}
 The old set of parameters is connected to the new one by a well defined function $\vec{\theta_{k}}=q(\vec{\theta_{k'}},u)$ 
(and of course $\vec{\theta_{k'}}=q'(\vec{\theta_{k}},u')$). We use {\it independent} proposals, so 
$\vec{\theta_{k}}=q(\vec{\theta_{k'}},u) = u$ and $\vec{\theta_{k'}}=q'(\vec{\theta_{k}},u')=u'$, thus, 
the Jacobian term in equation (\ref{eq:a'}) is unity. The algorithm spends most of the time iterating ``inside" the 
model that best describes the data. The RJMCMC method auto-penalizes high dimension models, also by 
taking into account the priors $p(\vec{\theta}_{k})$ of each model $k$. They serve as an Occam Factor integrated within the algorithm. 
Convergence is achieved if two main conditions are satisfied. First, the condition of {\it reversibility}, which is stated 
in a simple way: The proposal function must be invertible, meaning that we can jump from the proposed 
parameters back to the current parameters. And second, we must satisfy the {\em dimension matching} condition 
which in our case is always true since we use independent proposals in the acceptance ratio. After convergence 
has been achieved, a good approximation to the Bayes Factor is given by \cite{Littenberg2009, Bartolucci, Dellaportas2002, Cornish2007a} 
\begin{equation}
B_{\rm XY} = \frac{\rm \# \:of\: iterations\: in\: model\: X}{\rm \# \: of\: iterations\: in\: model\: 
Y}.
\label{eq:B}
\end{equation}

%
%
%
% OTHER APPROXIMATIONS
%
%
%

\subsection{\label{sec:otherapp}Other approximations}

While the RJMCMC method directly estimates the Bayes Factor, the other methods implemented in our work make an 
approximation to the evidence of each model. We implemented them as a cross-check for the RJMCMC 
method, but they also provide certain freedom of choice, depending on the nature of the problem, as well as the 
data available. 

The first approximations we consider are the Laplace approximations. The Laplace approximations
 perform a comparison between the volume of the models in the parameter space and the volume of the uncertainty 
 ellipsoid of the parameters \cite{bf1995}. This comparison is feasible if we assume that we work within a high Signal-to-Noise Ratio 
 (SNR) and therefore the posterior PDF is Gaussian near the maximum likelihood parameters, $\vec{\theta}_{\rm MAP}$. Another requirement is that a 
 sufficient large number of samples must have been collected. Then, the evidence in Eq.~(\ref{eq:evidence}) becomes:
\begin{equation}
\label{eq:L}
\pi_{\rm X}(\vec{\bf y}) \simeq  (2\pi)^{D_{\rm X}/2}  \left| \bf H \right| ^{1/2} \pi(\vec{\theta}_{\rm MAP, X},\: | \vec{\bf 
y}),
\end{equation}
where $D_{\rm X}$ is the number of dimensions of model X and ${ \bf H}$ is the Hessian matrix of the posterior . 
There are two main variations of the Laplace approximation. In the first one we make use of the Fisher Information 
Matrix $\bf F$, calculated at $\vec{\theta}_{\rm MAP}$, as an approximation to the expected covariance matrix \cite{Cornish2007a,Vallisneri2008}.
 Then, the evidence of the model becomes:
\begin{equation}
\label{eq:LF}
\pi_{\rm X}(\vec{\bf y}) \simeq (2\pi)^{D_{\rm X}/2} \left| \bf F \right| ^{-1/2} \pi(\vec{\theta}_{\rm MAP, X},\: | \vec{\bf y})
\end{equation}
Of course, the main limitations of this method are associated with the confidence we have on the calculation of the 
Fisher Matrix. Furthermore, as expected, the results appear to be poorer in comparison with the other methods 
as we move towards lower SNR areas. We can follow the notation of \cite{Cornish2007a} and call this 
particular approximation the Laplace-Fisher (LF) approximation. Another well-known variation is the Laplace-Metropolis (LM)
estimator of the marginal likelihood \cite{bf1995}. In this case, we use all necessary components for the calculation 
of the evidence from previous MCMC estimates. The parameters $\vec{\theta}_{\rm MAP}$ are extracted from the chains of a 
MCMC parameter estimation run for the particular model, while we use the weighted covariance matrix of the chains 
${\bf \Sigma}$, using a Minimum Volume Ellipsoid (MVE) or a Minimum Covariance Determinant estimator (MCD) \cite{rousseeuw}. 
The MVE method has been implemented and integrated in the LTPDA toolbox. In this case, the evidence of 
model X can be written as 
\begin{equation}
\label{eq:LM}
\pi_{\rm X}(\vec{\bf y}) \simeq (2\pi)^{D_{\rm X}/2}  \left| \bf \Sigma \right| ^{1/2} \pi(\vec{\theta}_{\rm MAP, X},\: | \vec{\bf 
y}).
\end{equation}
The LM method is considered to be a very reliable tool for the computation of the evidence of a model \cite{bf1995}. The third 
method we have used is the Schwarz-Bayes Information Criterion (SBIC) and is based on the following: After the assumption 
that the priors for each model follow a multivariate Gaussian PDF, the Schwarz criterion is defined as:
\begin{align}
\label{eq:SBIC}
S  \simeq  {\rm ln}(\pi(\vec{\theta}_{\rm MAP, X},\: | \vec{\bf y})) - {\rm ln}(\pi(\vec{\theta}_{\rm MAP, Y},\: | \vec{\bf y}))  \nonumber \\
- 1/2 (D_{\rm X} - D_{\rm Y}) {\rm ln}(n),
\end{align}
where $D_{\rm X}$ and $D_{\rm Y}$ are the dimensions of each model and $n$ is the number of samples 
in the data. It can be proven that if  $n \rightarrow \infty$, then $S \rightarrow B_{\rm XY}$ and the Schwarz 
criterion can be a good approximation to the Bayes factor. In fact, $n$ must be chosen carefully so that 
$n = N_{\rm eff}$, where $N_{\rm eff}$ is the number of effective samples in the data that represent the growth 
of the Hessian matrix of the log-likelihood \cite{bf1995}.

%
%
%
% IMPLEMENTATION
%
%
%

\subsection{Implementation} \label{sec:implementation}

If the system's transfer function is  ${\bf H}_{s}( \vec{\bf{\theta}})$ and 
we consider known injection signals, $\vec{\bf x}$, then the output of the system, $\vec{\bf y}$, can be expressed as: 
\begin{equation}
\label{eq:sys}
\vec{\bf y}={\bf H}_{s}( \vec{\bf{\theta}})\vec{\bf x}+\vec{\bf n},
\end{equation}
where $\vec{\bf n}$ is the overall instrument noise and $h(\vec{\theta}) = {\bf H}_{s}( \vec{\theta})\vec{\bf x}$ is the 
response or template of the system. The noise in the LTP experiment can be approximated with Gaussian noise since we work in a high 
SNR regime and we can safely consider that any dependence on the system parameters is negligible. Consequently, the likelihood 
for a model with parameters $\vec{\theta}$ can be written as
\begin{equation}
\label{eq:likelihood}
	\pi(\vec{\bf y}|\vec{\theta})=C\, {\rm exp}\big[ -\textstyle{1\over 2}\left< \vec{\bf y} - h(\vec{\theta}) | \vec{\bf y} - h(\vec{\theta}) \right> 
	\big],
\end{equation}
where the angular brackets denote the noise weighted inner product \cite{Barack}
\begin{equation}
\label{eq:inprod}
	<\vec{a}|\vec{b}> = 2 \int\limits_0^\infty \left[ \tilde{a}^\ast(f) \tilde{b}(f) + \tilde{a}(f) \tilde{b}^\ast(f) 
	\right]/S_n(f)\,,
\end{equation}
and where $S_n(f)$ is the power spectral density of the noise.
 For the proposal distribution appearing in the acceptance ratio of the Metropolis-Hastings algorithm a multivariate Gaussian distribution 
 is used. Although the choice of the proposal distribution should not affect the final estimated parameter PDFs,  
 it greatly affects the convergence speed of the chains to the target posterior distribution. The covariance 
 of the multivariate Gaussian is calculated beforehand or during the search phase using the Fisher Information Matrix (FIM) \cite{Vallisneri2008}.
 In the case of the RJMCMC method, the FIM is calculated once for each model before the execution of the algorithm. It is 
computed numerically or analytically  depending on the model. The LTP models are coded either in State-Space format \cite{Nofrariassymp,Hewitson2011,A.Grenadier}, or 
in pure analytical format in the acceleration domain \cite{accel}.
 %Each element of the FIM is defined as
%\begin{equation}
%\label{eq:FIM}
 %{\rm F}_{ij} = \left< \left(  \frac{\partial {\rm log} (\pi(\vec{\bf y}|\vec{\theta}))}{\partial \theta_i} \right)^T \left(  \frac{\partial {\rm log} (\pi(\vec{\bf y}|\vec{\theta}))}{\partial \theta_j} \right) \right> 
 %\Bigg|_{\vec{\theta}_0},
 %\end{equation} 
%and is calculated at a set of numerical values $\vec{\theta}_0$ of the parameters. The Cramer-Rao Bound (CRB), 
%as given by the components of the inverse of the FIM, provides an estimation of the expected errors in the parameters 

The first step in the analysis is to apply a Fourier transformation to the data and estimate the power spectral 
density of the noise $\vec{\bf n}$, i.e. $S_n(f)$. As described in \cite{Nofrarias2010}, the nature of the LTP experiment allows us to 
combine all the available information from different investigations. This means that we can combine the posterior
 distribution for each data set for each model. Then, we can apply MCMC analysis strategies to sample the posteriors and 
 make use of any of the criteria described in section \ref{sec:otherapp} or sample the joint parameter spaces with the  RJMCMC method 
  that was described in section \ref{sec:rj}. Technically, since the RJMCMC algorithm can be seen as a generalised MCMC 
  method, it performs a parameter search by simultaneously sampling the likelihood and consequently it maps the joint 
  posterior distribution and assigns PDFs to the parameters. The extra point is that we can estimate directly 
  the ratio of the evidences of the models and test the hypothesis made for the data.

%
%
%
% MODELING THE LTP
%
%
%

\section{Modeling the LTP experiment} \label{sec:ltpmodel}

\begin{figure*}[htb]
\begin{minipage}[b]{0.55\linewidth}
\includegraphics[width=\textwidth]{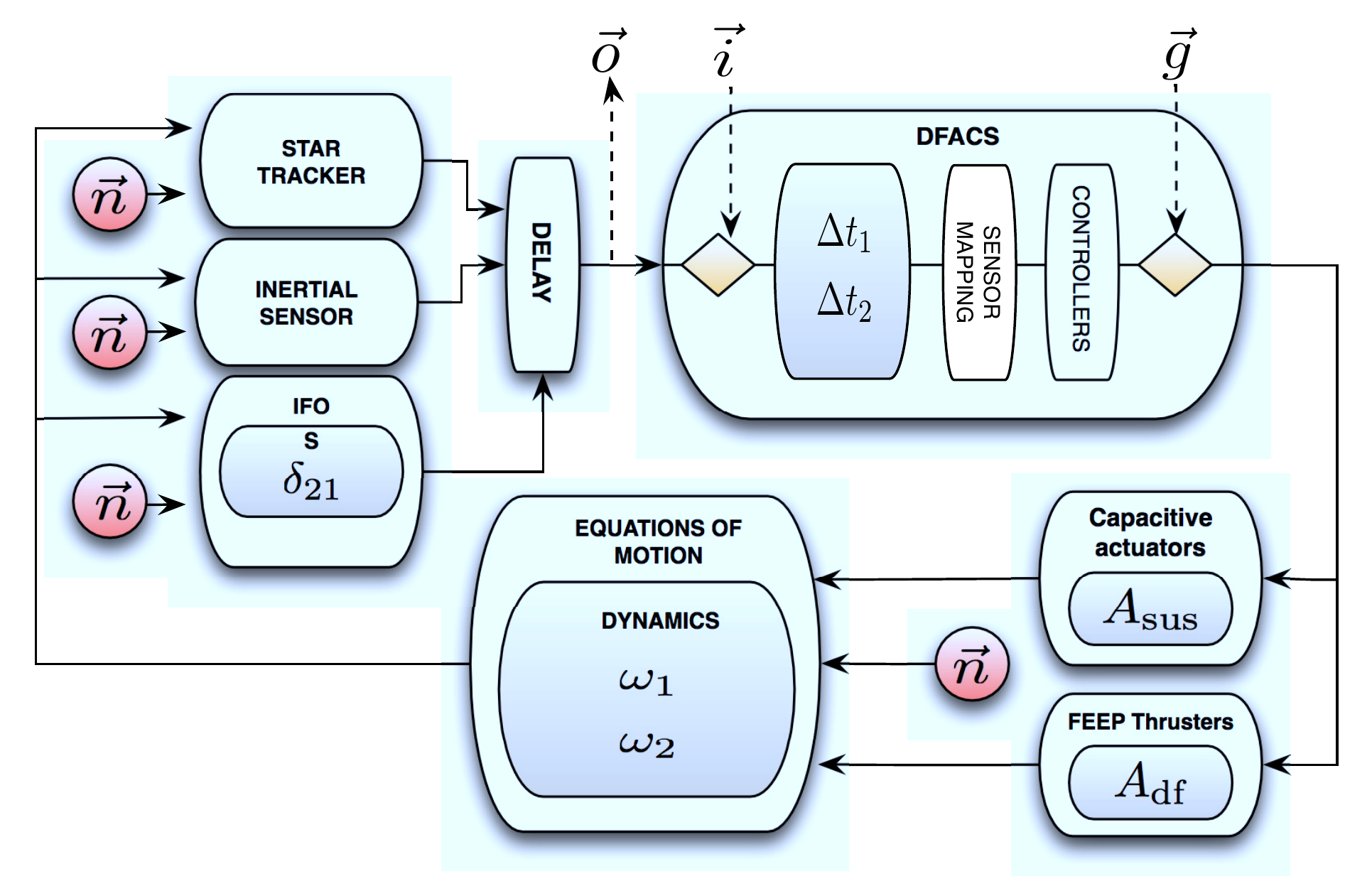}
\end{minipage}
\caption{Schematics of a simplified LTP State-Space Model (SSM). It is composed by smaller SSMs of the various 
subsystems, each one a standalone SSM, here represented with white boxes. The rhombi represent the
main injection inputs to the system. We use mainly the first injection port, which represents 
the interferometric inputs to the system, while the second rhombus stands for the capacitance actuators injection ports.
 The symbol $\vec{n}$ represents noise contributions from various sources and finally, the parameters to fit, for 
 the sake of convenience, are located in the gray boxes inside of the respective SSMs. The interferometer signals $\vec{i}$ are injected to 
 the controllers (DFACS), where the commanded forces are generated and applied through the capacitance 
 actuators and the thrusters of the space-craft. In the last two of experiments of \cite{TN3045} ``out of loop'' forces $\vec{g}$ are applied to the three bodies (TMs and SC) of the system. The resulting movement of the bodies is monitored by the Inertial Sensor, the Star Tracker and the interferometer. Here, the interferometer output is denoted as $\vec{o}$.}
\label{fig:LTP}
\end{figure*}

As a technology demonstrator of a space-based gravitational-wave observatory, 
the LPF mission will place two $2$~kg Test Masses ($\rm TM_1$ and $\rm TM_2$) in free fall. 
The goal is to estimate the residual differential acceleration between $\rm TM_1$ and $\rm TM_2$~\cite{Antonucci2011}. 
To minimize all external forces acting on $\rm TM_1$ along the x-axis, a 
 drag-free control loop~\cite{Fichter2005} has been designed. The coordinates of $\rm TM_2$ are controlled via electrostatic 
 actuators and the Spacecraft (SC) is controlled by micro-Newton thrusters. More specifically, the main components of the LPF mission are:
\newline 
\begin{itemize}

\item The Gravitational Reference Sensor (GRS)~\cite{Dolesi}. It consists of the test-masses and the vacuum chamber around 
them. It is mounted with two identical electrostatic actuators
to control the 6 degrees of freedom of each test-mass, and also to apply forces and torques to keep the test-masses in free fall.  

\item The Optical Metrology System~\cite{Heinzel} consists of the optical bench, its subsystems  
and the processing computer. It performs the sensitive optical measurements of the positions of the test-masses 
along the x-axis. For the simplified one-dimensional model version, we consider two interferometer inputs $\vec{i}$ and two 
outputs $\vec{o}$ of the system. The measured displacements are $x_1$ and $x_{12}$, the distance of ${\rm TM_1}$ to 
the SC and the distance between ${\rm TM_1}$ and ${\rm TM_2}$ respectively. 

\item The current design of the propulsion system for LPF is based on Cold Gas \cite{macnamara2013} thrusters. 
The main function of the thrusters is to maintain the reference test-masses in free fall conditions in the measured bandwidth. 

\item Finally, the Drag-Free and Attitude Control System (DFACS) calculates all forces and torques acting 
on the SC and test-masses and computes the commanded forces/response of the system in order to maintain the 
nominal free fall of $\rm TM_1$. 

\end{itemize} 

The LTP is a complicated instrument composed by a plethora of subsystems and coupled control 
loops. During the mission, there will be dedicated experiments with the aim of characterizing the different noise contributions 
and couplings. In the following section, the LTP models implemented for system identification 
experiments are described. We present simulations of the planned system identification experiments and use all techniques 
described to perform model selection.

%
%
%
% THE DYNAMICS
%
%
%

\subsection{The LTP system dynamics} \label{sec:sysdyn}

For the purposes of this work we consider a simplified version of the LTP operating in the so-called main 
science mode and we confine the system dynamics to the one-dimensional case. There are two approaches in modeling the dynamics of the three-body system:
The first one, shown in Fig. \ref{fig:LTP}, is to fully describe the dynamics together with the controllers in a state-space format (SSM). In this case, it is very convenient to 
represent the model as a closed loop system, due to closed-loop dynamics and controllers of the instrument. The white boxes represent
standalone SSMs, all together {\em assembled} to form the final LTP experiment~\cite{Nofrariassymp}. Each submodule 
has injection $u(t)$ and output $y(t)$ ports and can be represented as:  
\begin{gather}
  \label{eq:ssm}
  \dot{x}(t) = A\times x(t)+B\times u(t),  \nonumber \\
  y(t) = C\times x(t)+D \times u(t),
\end{gather}
where $x(t)$ are the {\em states} of the system, A is the state matrix, B is the input matrix, C is the output 
matrix, and D is the feed-through matrix \cite{Nofrariassymp}. This form of implementation has numerous advantages,
like modularity and flexibility in modeling the LPF mission. For example, the user is able to 
combine any given SSM module and use a custom noise model for each particular subsystem. 

The second approach is to represent the dynamics in an analytical way in the acceleration domain \cite{accel} which is described in section \ref{sec:polezero}. 
A more in-depth investigation on the analysis in acceleration domain and its practical advantages is to appear soon.

For the purposes of this paper, we simulate the experiments as explained in \cite{Nofrarias2010,Congedoprd,Nofrariassymp,Nofrarias:2011pq,TN3045}. Then, the complete
 system is controlled by optical readouts that measure the positions of the
 Test Masses, $x_1$ and $x_{12}$, where the $x$-axis is defined by the line joining them. 
 In order to perform parameter estimation we inject sinusoidal signals to each interferometric channel alternately, or direct forces to the three bodies of the system (test-masses and SC), and measure the response of the system. For this type of experiments, we can define the following time-series arranged as vectors:
\begin{equation}
\label{eq:signals}
\vec{o}=
\begin{pmatrix}
 o_{x_1} \\
 o_{x_{12}} 
\end{pmatrix}, \qquad 
\vec{i}=
\begin{pmatrix}
 i_{x_1} \\
 i_{x_{12}} 
\end{pmatrix}, \qquad 
\vec{n}=
\begin{pmatrix}
 n_{x_1} \\
 n_{x_{12}} 
\end{pmatrix}, 
\end{equation} 
\begin{equation}
\vec{g}=
\begin{pmatrix}
 g_{1} - g_{SC} \\
 g_2 - g_1
\end{pmatrix},
\end{equation}
where $\vec{o}$ is the measured signals vector, $\vec{i}$ is the interferometer injection signals vector, $\vec{g}$ is the commanded forces signals vector, and $\vec{n}$ is the overall 
noise for both channels. Here $g_1$ and $g_2$ are the commanded forces applied to $TM_1$ and $TM_2$ respectively while $g_{SC}$ is the force applied to the SC. The noise contributions of each LTP subsystem are denoted in Fig. \ref{fig:LTP} 
as $\vec{n}$. It can be instrumental or read-out noise, or it may originate from external sources (like solar radiation). 

For the sake of simplicity, we use the one-dimensional 
version of the models. The set of parameters that we recover from 
the system is $\{ \omega_1, \omega_2, \Delta t_{1}, \Delta t_{2}, A_{\rm df}, A_{\rm sus}, \delta_{21}\}$: 
 $\omega_1$ and $\omega_2$ stand for the electrostatic stiffness of the test-masses 
to the surroundings, $\Delta t_{1}$ and $\Delta t_{2}$ are time delays, $A_{\rm df}$ and $A_{\rm sus}$
the capacitance and thruster gains respectively, and $\delta_{21}$ denotes 
the cross-coupling between the two interferometric channels.

As described in \cite{TN3045,Nofrarias:2011pq}, during the planned mission experiments, sinusoidal signals 
of different frequencies are going to be injected to the 
 control loop, simulating displacements of the test masses. Besides the LTP dynamics system, the data analysis team has 
 modeled noise sources for the various subsystems of the LPF mission based on theoretical predictions or characterization 
 of each element from on-ground test campaigns. The main noise contributions come from the thrusters, the 
 interferometric readouts, and the capacitance actuators \cite{lpf_nb}. 
 We can simulate synthetic noise for any particular experiment and perform parameter estimation exercises in order to
  test the system identification algorithm's readiness. A MCMC search of the parameter space for such a simulated
   run returns satisfactory results as shown in Table \ref{tab:mcmc}. 

\begin{table}[h]
\begin{center}
\caption{MCMC parameter estimation results from simulated experiments. The experiments being analyzed here, 
are the two first described in \cite{Nofrarias2010} and  \cite{TN3045}.}
\label{tab:mcmc}
\begin{tabular}{l c r c l} 
\hline
\hline
Parameter & Real value & \multicolumn{3}{c}{Estimated $\pm \sigma$}  \\
\hline
{\scriptsize $\omega_1$ } & \footnotesize $1.3\times10^{-6}$ & \footnotesize $(-1.2999$ & $\pm$ & $ 0.0002)\times10^{-6}$ \\
{\scriptsize $\omega_2$ } &  \footnotesize$1.9\times10^{-6}$ &\footnotesize $(-1.8999 $& $\pm$ &$ 0.0002)\times10^{-6}$ \\
{\scriptsize $A_{\rm df}$ } & \footnotesize  0.82 & $ 0.8201 $& $\pm$ &$0.00025 $ \\
{\scriptsize $A_{\rm sus}$ } &  \footnotesize1.08  & \footnotesize $1.080004 $& $\pm$ &$ 3 \times10^{-6}$  \\
{\scriptsize $\Delta t_{1}$ }& \footnotesize0.2 & $\footnotesize 0.20020  $& $\pm$ &$ 5\times10^{-5}$ \\
{\scriptsize $\Delta t_{2}$  } & \footnotesize0.2 & $\footnotesize 0.2003  $& $\pm$ &$ 2\times10^{-4}$ \\
{\scriptsize $\delta_{21}$  } & \footnotesize0.0004 & $\footnotesize 0.0004001 $& $\pm$ &$ 1\times10^{-7}$ \\
\hline
\hline
\end{tabular}
\end{center}
\end{table}

By using the LTPDA machinery, synthetic noise can be generated and 
 the response of a known system can be simulated. For the following study cases, the LTP default 
 parameters were always set to the same values given in Table~\ref{tab:mcmc}. It should be noted that there is no 
reason to have any strong prior belief about the true values of the parameters. 
Therefore, it is reasonable to assume uniform priors for all the tests. 

%
%
%
% APPLICATIONS
%
%
%

\subsection{Application to a simplified LTP model} \label{sec:guidanceinv}
In order to demonstrate a first application to the LTP, we can investigate a model selection case that was
first encountered during a data analysis exercise \cite{Nofrarias:2011pq}. This type of exercises are organized by the DA team with the aim of testing the developed 
tools in more realistic scenarios. 
In this case, the ``true'' values of the parameters of the system to identify, were totally unknown.
In that situation, the data analysis team was able to perform parameter estimation, but it was noted that the fit was improved
 when two extra parameters were introduced. These parameters 
 are the guidance delays, $\Delta t_{1}$ and  $\Delta t_{2}$. They are 
 simply time delays to the application of the guidance signals (see Fig. \ref{fig:LTP}), due to 
 operations of the Data Management Unit (DMU), which is the on-board computer controlling the LTP experiment. The final 
 robustness of the fit indicates that the new parameters substantially improve the model.  
 
This problem can be better studied by reducing it to a model selection problem and can 
be tested with synthetic data-sets. The simulated data source is a LTP system with the default characteristics given in Table~\ref{tab:mcmc} with the exception of
 the delays, which were set to: $\Delta t_{1} = \Delta t_{2} = 0$. With this 
 configuration, we assume a true system where the application signals are applied instantaneously. For our 
 first try-out we injected ``fake'' interferometric displacements, while for the second investigation we 
 used the same structure of injections but with much lower Signal to Noise Ratio (SNR $\sim 5$). 

\begin{figure}[!ht]
\begin{center}
\includegraphics[width=20.5pc]{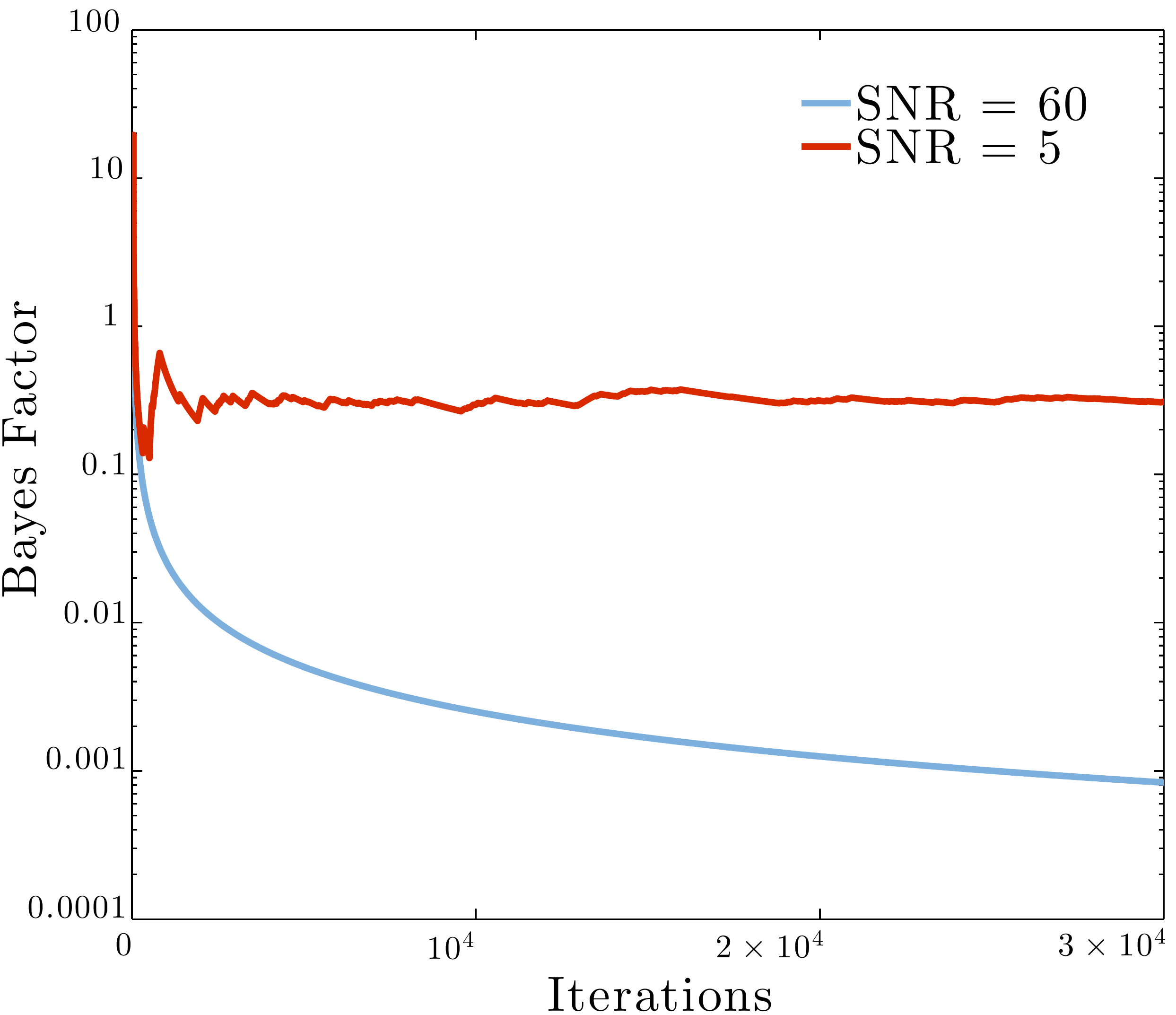}
\caption{ (Color online) First $3\times10^{4}$ iteration of the RJMCMC output when comparing a seven and a 
five-dimensional LTP model (models X and Y respectively). Since the models are not competitive when the ${\rm SNR}=60$,
 the blue line tends asymptotically to zero.}
\label{fig:Bxy_del_inv}
\end{center}
\end{figure}

In order to determine the importance of 
the two extra parameters, we have to verify 
which model describes better the particular data-set: The seven parameter model X with parameters 
$\vec{\theta}_{\rm X} = \{ \omega_1, \omega_2, \Delta t_{1}, \Delta t_{2}, A_{\rm df}, A_{\rm sus}, \delta_{21}\}$,
 or the five parameter model Y with parameters   
 $\vec{\theta}_{\rm Y} = \{ \omega_1, \omega_2, A_{\rm df}, A_{\rm sus}, \delta_{21}\}$. 
 While both models, X and Y, are capable of explaining the observations, we expect the simpler 
 model Y to be more favorable from a RJMCMC output, since the extra parameters $\Delta t_1$ and $\Delta t_2$ are not significant for the data. 
 The evolution of the Bayes factor for such an investigation for two different levels of SNR is shown in 
 Fig.~\ref{fig:Bxy_del_inv}. 
\begin{table}[h]
\begin{center}
\caption{Results for the Guidance delay investigation with the low SNR experiment. See text for details.} 
\label{tab:results_LTP_delay}
\begin{tabular}{C{3cm} C{3cm}}
\hline
\hline
Method &   $B_{\rm XY}$  \\
 \hline
RJMCMC  &  0.309  \\
LF  & 0.124  \\
LM  & 0.078 \\
SBIC  & 0.768  \\
\hline
\hline
\end{tabular}
\end{center}
\end{table}

 The results we obtain for all the approximations verify that the simpler model Y is much more 
 probable than the more complicated model X. 
 For the particular experiments proposed in \cite{TN3045} where the SNR is high, 
 the Bayes factor computed tends to zero. In fact, for the case of the RJMCMC method, there is no single iteration 
 ``inside'' the more complex model. This changes dramatically depending on the nature of the problem and, of course, 
 as we show in section \ref{sec:BFvsSNR}, on the SNR. In Table~\ref{tab:results_LTP_delay} we present only the low-SNR experiment, for the 
sake of comparison between the methods. Each method seems to favor the simpler model but they are not in total agreement between them. 
This is to be expected, as the SNR of this investigation is very low and the approximations of the evidence become more sensitive. 
For this particular case of the injections, the models are not competitive and therefore, 
the resulting estimated Bayes Factor is extremely small. For a direct comparison, the RJMCMC algorithm requires more than $10^{8}$-$10^{9}$ iterations.

%
%
%
% A MORE REALISTIC APPLICATION
%
%
%

\subsection{More realistic applications} \label{sec:polezero}
\begin{figure*}[ht]
\begin{minipage}[t]{0.45\linewidth}
\centering
\hspace*{-0.8cm}
\vspace{0pt} 
\includegraphics[width=20.3pc]{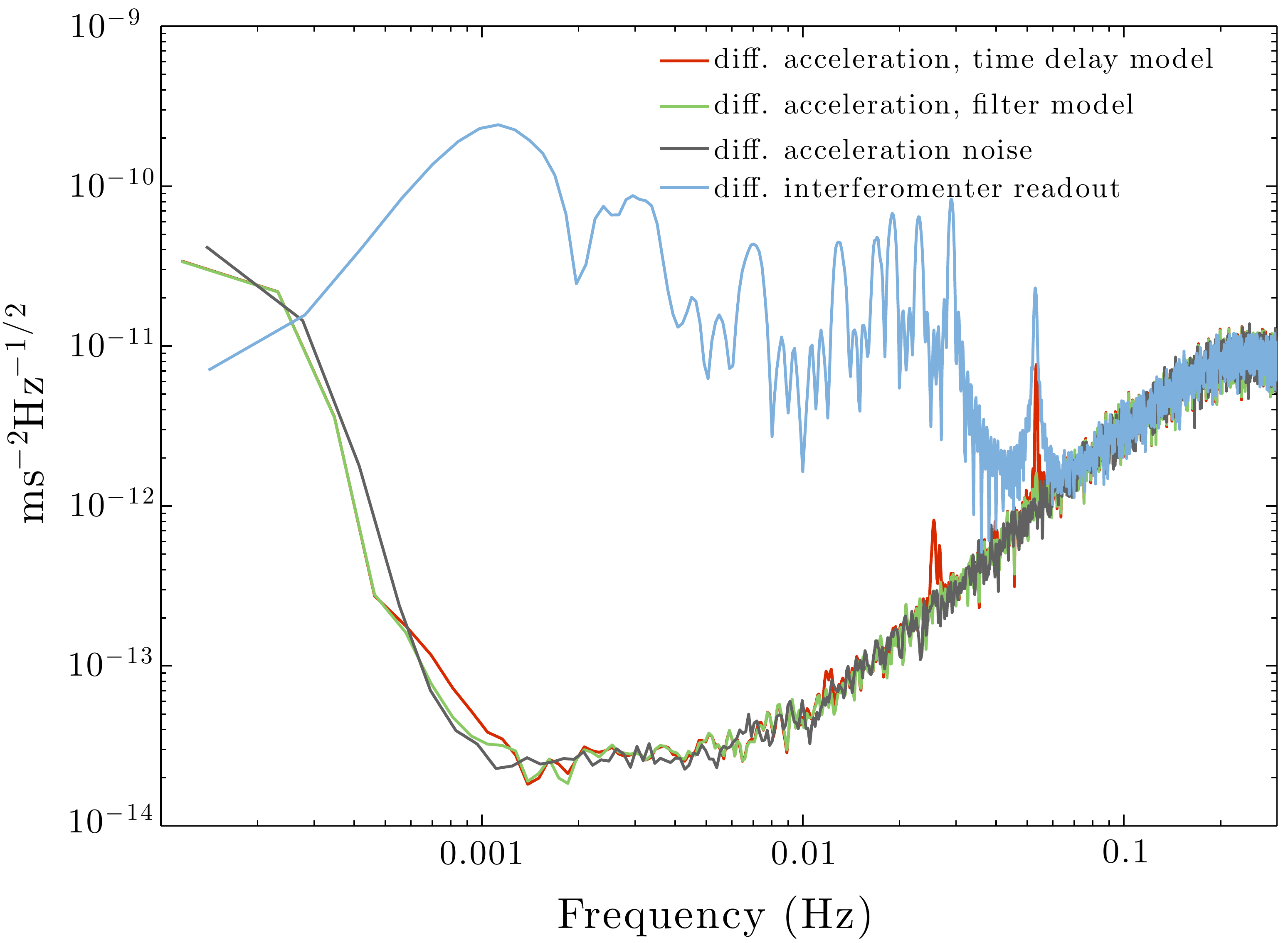}
\caption{(Color online) Power spectra of simulated differential acceleration between the two test-masses. The gray curve represents the reference noise measurement, while the light blue curve is the differential interferometer read-out. The value of the computed Bayes Factor can be confirmed with the comparison of the equivalent estimated residual acceleration for each model. The differential interferometer read-out is also plotted for comparison.}
\label{fig:poledelaynew}
\end{minipage}
\hspace{0.2cm}
\begin{minipage}[t]{0.45\linewidth}
\centering
\includegraphics[width=20pc]{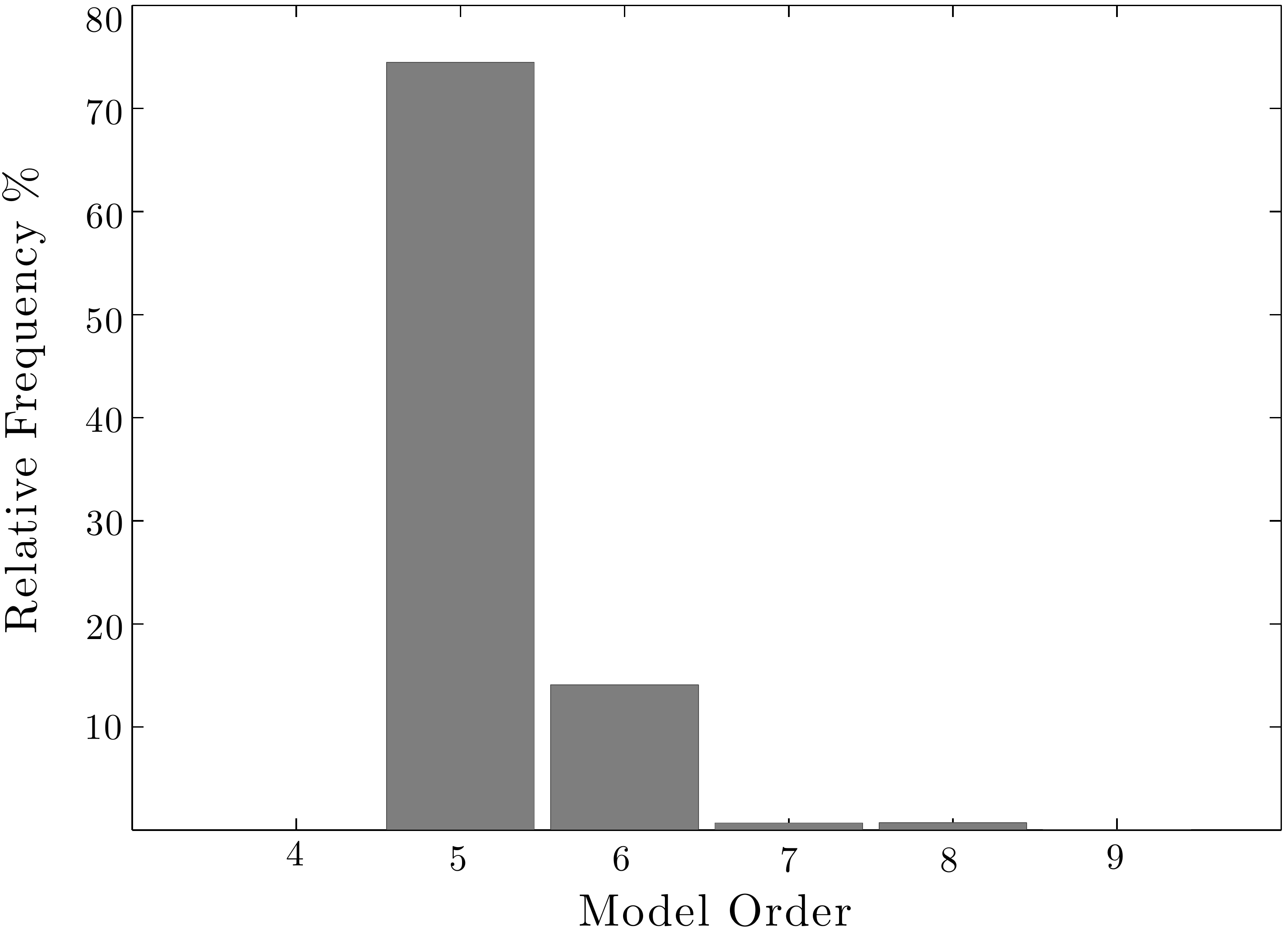}
\caption{A RJMCMC run on a set of nested LTP models. There is a clear preference for the five-dimensional model for the given data-set. The data were produced with a ``perfect'' model where the two respective actuators were identical.}
\label{fig:order_hist}
\end{minipage}
\end{figure*}

During the last years a series of data analysis exercises have been performed. These so called Operational Exercises are organized by the LPF Data Analysis team with the aim
of training the scientists and engineers for the upcoming flight operations. Data generated by the ESA simulator are used and real-life shifts and duties are rehearsed. The ESA simulator for the LPF is an industry developed software that is considered the most accurate simulator to date. The 
scientific and organizational results are valuable since they reproduce situations and data analysis challenges that are expected to occur during flight operations. 

An outcome of these data challenges was the realization of the necessity to change or manipulate the model of the dynamics of the system shortly after receiving telemetry from the satellite, thus facing once again model selection problems. Expressing the 
dynamics in the acceleration domain \cite{accel}, one could write a simplified model for the differential acceleration, like the one below:
\begin{align}
a_{12} = & \left[ \frac{d^2}{dt^2} + \omega_2^2 \right] x_{12} + (\omega_{2}^2 -\omega_{1}^2)  x_1 \nonumber \\ 
             & - A F_{2} + A F_{1} 
\label{eq:a12}
\end{align}
where $a_{12}$ is the differential acceleration and $F_1$ and $F_2$ denote the applied forces on the first and second test-masses respectively. The real motion $\vec{x}$ of the test-masses can be approximated by the delayed interferometer read-outs:
\begin{equation}
\label{eq:truemotion}
\vec{x}=
\begin{pmatrix}
 x_{1} \\
 x_{12} 
\end{pmatrix}
=
\begin{pmatrix}
 o_{1}(t - \tau_{ifo}) \\
 o_{12} (t - \tau_{ifo})
\end{pmatrix}
\end{equation}
 The parameters appearing in Eq. (\ref{eq:a12}) and (\ref{eq:truemotion})  are the stiffnesses of the two test-masses $\omega_1$ and $\omega_2$, the interferometer read-out delay $\tau_{ifo}$, and the gains of the capacitance actuators $A_{sus}$ (here represented as $A$ for simplicity). Note that for this first approximation we have assumed identical actuators $A$ for both test-masses. The commanded forces $g_1$ and $g_2$ are available as telemetry, but the {\em real} applied forces on the three bodies are
to be determined by the measurements and the analysis itself. For example, in the real data-stream there might be additional delays or even filtering of the applied forces
coming from the controllers, so in reality a gain $A$ might be proven to be frequency dependent: $A(f)$. This situation will appear in the received telemetry and we need the means to disentangle those two physical
effects in a quantitative way.

For the particular simulated data-set, the model was not able to remove all the injected signals and even for the simple case of Eq. (\ref{eq:a12}), this led to a biased estimate for the parameters.
Apart from a simple time delay on the commanded signals, there might be another process that causes a difference between the commanded $g_1$ and $g_2$ and the actual applied forces $F_1$ and $F_2$.
For a first approximation of such a process,  we assume a single real pole filter, filtering the time-series of the applied force on the second test-mass. For this investigation we can propose two models where $A F_i (t) = A g_i (t-\tau_{C})$ and $A F_i  =  A (\frac{f_o}{f - jf_o}) g_i$ respectively. Here $\tau_{C}$ denotes the actuators time delay. In the end we can apply the model selection methods to these two models: the first one, X, where the applied forces are time delayed, and the second one, Y, where the forces are frequency dependent (filtered by a single real pole filter). The calculated Bayes Factor between those two models is:
\begin{equation}
{\rm BF_{XY}} = \frac{\pi_{\rm X}(\vec{\bf{y}})}{\pi_{\rm Y}(\vec{\bf{y}})} = 9.9478 \times 10^{-11} ,
\label{eq:bfpoledelay}
\end{equation}
 clearly indicating that the most probable process on the forces is the one described by model Y. The real pole was estimated to be
$f_o = 1.963\pm0.001$ Hz and was confirmed for the complete set of system identification experiments. The estimated acceleration residuals for both models can be seen in Fig. \ref{fig:poledelaynew}.

The same analysis strategy applies of course to more complicated versions of the analytical models, where the number of parameter increases with the more terms of the equation are added, as happens in the following more realistic case. As already
described in section \ref{sec:ltpmodel}, the two test masses are controlled with identical actuators. In reality, a small misbalance between the electrostatic actuators surrounding each test mass might be present. If we introduce this ``asymmetry'' to the system, immediately for the simple case of Eq. (\ref{eq:a12}), we can increase the dimensionality of the model by four parameters:
\begin{align}
a_{12} = & \left[ \frac{d^2}{dt^2} + \omega_2^2 \right] x_{12} + (\omega_{2}^2 -\omega_{1}^2)  x_1 \nonumber \\ 
             & - A_{2} ( \frac{f_{o2}}{f - jf_{o2}})  g_{2}(t-\tau_{C2})  \nonumber \\
             & +  A_{1} ( \frac{f_{o1}}{f - jf_{o1}}) g_{1}(t-\tau_{C1}) .
\label{eq:a12+}
\end{align}
 These parameters are gains, delays and filters that are different for the actuator of each test mass. Theoretically the highest in dimensions model of Eq.(\ref{eq:a12+}) can describe the observations, but the problem to solve appears to be over-parametrized. A solution is to generate a set of nested models under the highest in dimensions of Eq. (\ref{eq:a12+}) and apply the RJMCMC algorithm. The result of such a run is shown in Fig. \ref{fig:order_hist} and it reveals the most favorable model and consequently the underlying procedure that describe best the physical system. For the particular simulation we can verify that the five-dimensional model is the best, concluding that no ``asymmetry'' in the hardware of the LTP is present. This changed in the following Data Challenge, where the Bayes factor between the simple model of Eq. (\ref{eq:a12}) and a six-parameter asymmetric one, is greater than $10^6$, clearly supporting the correct higher dimension model where $A_{1} \neq A_{2}$.
 \begin{figure*}[ht]
\begin{minipage}[t]{0.45\linewidth}
\centering
\hspace*{-0.8cm}
\vspace{0pt} 
\includegraphics[width=21pc]{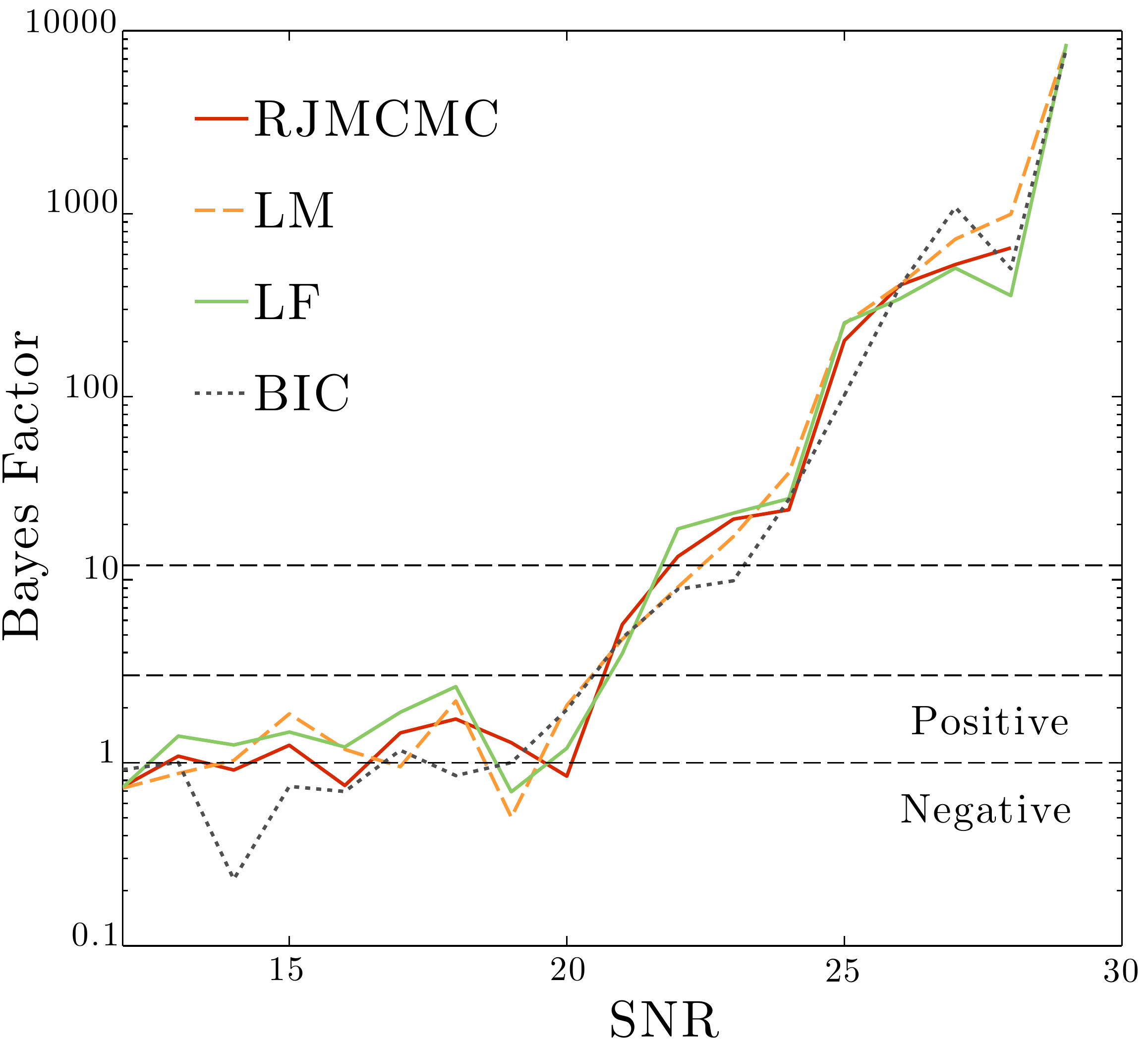}
\caption{(Color online) The Bayes Factor as a function of SNR computed using different methods.}
\label{fig:b_snr}
\end{minipage}
\hspace{0.2cm}
\begin{minipage}[t]{0.45\linewidth}
\centering
\includegraphics[width=21pc]{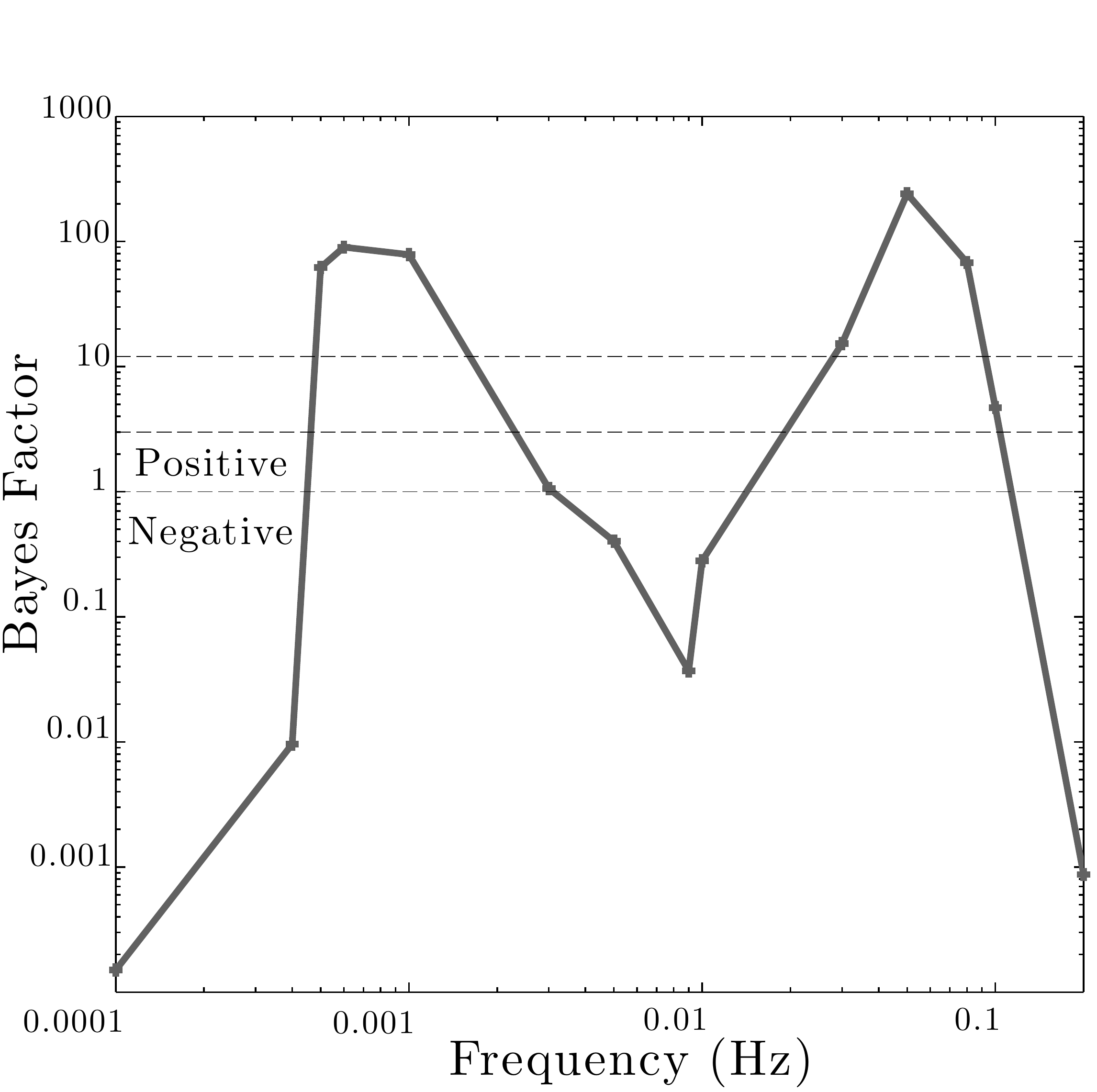}
\caption{(Color online) The Bayes Factor as a function of the injection frequencies in the first interferometric channel. The computed evidence of model X is stronger when the sinusoidal injection signals have a frequencies around $f_{i_{x_1}}\simeq 0.0006$ Hz and $f_{i_{x_1}} \simeq 0.05$ Hz respectively.}
\label{fig:B_freq_snr}
\end{minipage}
\end{figure*}

%
%
%
% THE BAYES FACTOR FOR EXPERIMENT OPTIMIZATION
%
%
%

\section{Using the Bayes factor for experiment optimization}
\label{sec:BFvsExp}

In this last section we explore the capability of the implemented model selection framework, not only as 
a set of tools for data analysis purposes, but also as a way to evaluate the efficiency of the experiments we are planning 
to run in the satellite. As we are going to show, by comparing different models under different 
input signal conditions, we can safely determine the best range of parameters that define our experiments, or
verify the injection frequencies that maximize the information extracted from the system. 

%
%
%
% THE BAYES FACTOR AS A FUNCTION OF SNR
%
%
%

\subsection{The Bayes Factor as function of the SNR}\label{sec:BFvsSNR}

It has been shown~\cite{Cornish2007a,Veitch} that there is a dependence of the Bayes Factor output on 
the SNR regime of the investigations. This, of course, holds true in the case of 
the LTP as it can be seen in Fig.~\ref{fig:b_snr}.  This figure was created by 
simulating LTP experiments for each value of the SNR, while the injection signals were single 
frequency ($f = 0.01 {\rm Hz}$) sinusoidal inputs into the system. The LTP models under comparison were 
quite similar with the exception of a different realization of the response model of the 
thrusters.  
It is clear that above the critical value of  the $\rm SNR = 21$
 the results obtained with the different techniques are consistent and in good agreement. Below that value of the 
 SNR we cannot make clear decisions about the competing models, as the wrong model is showing preference,
  or we poorly approximate the Bayes Factor.
  
Although this SNR limit varies, as expected, depending 
 on the type of investigation and model, the current result is already providing an 
 estimation of the required power of the injection signals that we need to consider in the LTP 
 experiment. This information and the method used here will be of 
 particular interest for the design of in-flight experiments for the LPF mission.
 
%
%
%
% THE BAYES FACTOR AS A FUNCTION OF THE EXPERIMENT DESIGN
%
%
%

\subsection{The Bayes Factor as function of the injection frequencies}\label{sec:BFvsF}
   
 Furthermore, for system identification experiments, as in the case of the LTP, the computed 
 evidence of a model depends on the design of the experiment itself. The 
 information obtained from the system differs depending on the injection 
 frequencies. An interesting study is to explore in detail this relation. 
 A four- (X) and a five-dimensional (Y) models are examined, given 
 different injection frequencies. More precisely, since the difference between the models is the cross-coupling $\delta_{21}$ as
 shown in Fig. \ref{fig:LTP}, which describes the signal leakage from the first to the differential 
 interferometric channel, we examine the Bayes Factor given different injection 
 frequencies to the first channel, while keeping constant the injection to the 
 differential channel ($f_{i_{x12}} = 0.2$~Hz). The SNR of this experiment 
 is kept at the ``low'' value of 28.
 
 The data generation model is mounted with a ``perfect'' interferometer ($\delta_{21}=0$) and model X is the 
 same as the one used to produce the data, while model Y is the one with the extra parameter $\delta_{21}$. 
 The expected outcome of this exploration is that if the system is more sensitive to the $\delta_{21}$ parameter
 at some particular frequencies, we must detect an increase in the Bayes Factor which underlines a more clear decision 
 towards the correct model.

 In Fig.~\ref{fig:B_freq_snr} we can see the corresponding Bayes Factor versus the injected frequencies to 
 the first channel. Given the low SNR of the investigation, while model X 
  should be more favorable, a preference for the more complex model Y is shown 
  for a certain set of frequencies. This result is a clear indicator of the set 
  of preferred frequencies that can be injected to the system for its 
  characterization given the current configuration and SNR. Indeed, injections 
  around $f \simeq 0.0006$ Hz and $f \simeq 0.05$ Hz promote the identification of 
  the correct model, while injections at both the high and low frequency limit, 
  together with $f\simeq 0.01$ Hz, may induce the analysis into an error. This 
  frequency dependence must be associated to the sensitivity of the experiment to a given parameter, the 
  parameter $\delta_{21}$ in this particular case. This observed dependency, when 
  considering a more realistic model, will be of particular interest in the selection of injection signals for the experiments
  to be run in-flight. 

%
%
%
% CONCLUSIONS
%
%
%
 
\section{Conclusions}\label{sec:conclusions}

We have implemented three different methods to compare competing models of the LTP experiment on-board the LPF 
mission: The RJMCMC algorithm, the Laplace approximations, and the Bayes Information Criterion.
The results from each method seem to be in agreement, but the output strongly 
  depends on the expected SNR and of course on the models under investigation. Considering the LPF mission planned experiments, 
  the SNR is high enough to safely use any of all the available techniques, but probably the
   most computationally demanding methods will be used for off-line analysis to confirm our first computations.
   
The RJMCMC algorithm (together with the Laplace methods and the Bayes Information
 Criterion) employed in this work has been integrated in the LTPDA toolbox as part of the LPF data analysis software. 
 The RJMCMC algorithm is by far the most computationally costly, 
 but at the same time it is the more suitable one when we compare more than two nested models or we work with inputs with low SNR.  
The Laplace-Metropolis and the Laplace-Fisher methods are reliable when we work in the high SNR regime, 
but they also require significant computing time, specially when one has to use outlier detection methods to estimate 
the weighted covariance matrix. On the other hand, the Laplace-Fisher approximation is limited by 
the use of the Fisher Information Matrix, which for the case of LTP state space models 
is computed numerically. 

 Moreover, an attempt to associate the output of the aforementioned methods with the actual system identification experiment has been made. We have used 
  different experiment setups to demonstrate that the Bayes Factor 
  depends not only on the SNR, but also on the injection frequencies to the system.
  
The developed algorithms were successfully applied to model selection problems for the LPF data analysis for the first time. 
Two different cases of LTP model selection problems have been investigated over data-sets that were produced by both the 
LTPDA and the ESA simulator. For the first case, we have considered an easy case of five- and seven-dimensional state-space models, where the importance of the 
 extra two parameters was examined. These two extra parameters are time delays caused by the LPF hardware and 
 they can be characterized as essential parameters of the model. For the second case, we explored the most suitable dimensionality of analytic models. There, the simplest model that described efficiently the observations 
 was recovered, excluding the more complicated ones that caused over-fitting issues. This type of analysis is expected to be performed  
 during operations due to the broad spectrum of possible applications, like identifying external disturbances that result into forces applied to the three-body
 system.
\newline

%
%
%
% ACKNOWLEDGEMENTS
%
%
%

\acknowledgments  We acknowledge support from contracts AYA-2010-15709 (MICINN) and 2009-SGR-935 (AGAUR). 
NK acknowledges support from the FI PhD program of the AGAUR (Generalitat de Catalunya). 
MN acknowledges a JAE-doc grant from CSIC and support from the EU Marie Curie CIG 322288. 
GC acknowledges support from the Beecroft Institute for Particle Astrophysics and Cosmology.

%
%
%
% REFERENCES
%
%
%

%\bibliography{library}{}
%\bibliographystyle{apsrev}

\end{document}